\documentclass[prl,twocolumn,floats,aps,epsfig,nofootinbib,amssymb]{revtex4}
\usepackage{subfigure}
\usepackage{graphicx}
\usepackage{cancel}
\usepackage{amssymb}
\usepackage{textcomp}
\usepackage{amsmath}
\usepackage{bm}
\usepackage{times}
\usepackage{epsfig}
\usepackage{color}
\def\beq{\begin{equation}}
\def\eeq{\end{equation}}
\def\bea{\begin{eqnarray}}
\def\eea{\end{eqnarray}}

\begin{document}

\title{\Large {\textrm{Low Energy Supersymmetry with Baryon and Lepton Number Gauged}}}
\bigskip
\author{Pavel Fileviez P{\'e}rez}
\address{Phenomenology Institute, Department of Physics, University of Wisconsin-Madison, WI 53706, USA}
\author{Mark B. Wise} 
\address{
California Institute of Technology,  1200 East California Blvd., Pasadena, CA 91125, USA}
\date{\today}

\begin{abstract}
We investigate the spontaneous breaking of the Baryon (B) and Lepton (L) number at the TeV scale in supersymmetric models. 
A simple extension of the minimal supersymmetric standard model where B and L are spontaneously broken local gauge 
symmetries is proposed. The B and L symmetry breaking scales are defined by the supersymmetry breaking scale.  
By gauging B and L we understand the absence of the baryon and lepton number violating interactions of dimension four and five in the MSSM. 
Furthermore we show that even though these symmetries are spontaneously broken there are no 
dangerous operators mediating proton decay. We discuss the main properties of the spectrum, the possible baryon number violating decays 
and the implications for the dark matter candidates. In this model one can have lepton number violating signals from the decays 
of the right-handed neutrinos and baryon number violating signals from the decays of squarks and gauginos 
without conflict with the bounds coming from proton decay, $n-\bar{n}$ oscillations and dinucleon decays.
\end{abstract}

\maketitle
{\section{I. Introduction}}
Experimental data are consistent with baryon number (B) conservation and lepton number (L) conservation. 
In neutrino experiments we have observed the violation of the individual lepton numbers ${\rm L}_{e,\mu,\tau}$ 
but not of the total lepton number ${\rm  }L={\rm L}_{e}+{\rm L}_{\mu}+{\rm L}_{\tau}$. It is interesting to 
explore the possibility that the observed B and/or L conservation has its origin in the principle of gauge invariance 
and construct models where B and L are spontaneously broken gauge symmetries. To gauge B and/or L additional 
fermions beyond those in the minimal standard model must be added to cancel anomalies. Solutions to the anomaly 
constraint equations were found in Ref.~\cite{Foot:1989ts,Carone,BL}.  

The authors in Ref.~\cite{Carone} explored models where baryon number is gauged with the anomalies canceled 
by adding a fourth generation of quarks and leptons. Since three generations have been observed, and we do not 
understand why there should be only three,  we view this way of canceling anomalies as less arbitrary than 
the other possibilities for canceling anomalies that introduce fermions with quantum numbers unrelated to 
those of the observed standard model fermions. Recently, we constructed two explicit models where both B and L 
are spontaneously broken local gauge symmetries~\cite{BL}. In these models B and L are on the 
same footing and the anomalies are cancelled by adding a single new fermionic generation. There is a natural 
suppression of flavour violation in the quark and leptonic sectors since the gauge symmetries and particle content 
forbid tree level flavor changing neutral currents involving the quarks or charged leptons. Also there is a dark matter 
candidate that is automatically stable. In these models the symmetry breaking scale for the $U(1)_B$ and $U(1)_L$
 symmetries are not necessarily related to the weak scale however we explored some of their phenomenology with that assumption.

In the standard model operators that violate baryon number (schematically $qqql$) do not occur until dimension six and experimental constraints on the nucleon decay rate imply that the mass scale that suppresses them, $\Lambda$  must satisfy, $\Lambda > 10^{15}~{\rm GeV}$.  Hence the observed conservation of baryon number is explained if there is no new physics below this scale, {{\it \i.e.}, a desert}. However in models where baryon number is gauged the observed conservation of baryon number  can be understood, even if there is  new physics at scales much lower than $10^{15}~{\rm GeV}$, since without spontaneous symmetry breaking operators that violate B  are forbidden and (depending on the charges of the fields that break baryon number) the spontaneous breaking of baryon number may not induce operators that cause observable proton decay.

Supersymmetry (SUSY), softly broken at the weak scale, solves the hierarchy problem. Today, the minimal supersymmetric extension 
of the standard model (MSSM) is considered one of the most appealing scenarios for physics beyond the Standard Model (SM). 
For a review on supersymmetric models, see Ref.~\cite{Manuel}. One of the open issues for these models is the presence of renormalizable and dimension five operators that violate  baryon and lepton number.  These can be forbidden by gauging a linear combination of B and L~\cite{Lee:2010hf} and it is interesting to consider extending the work in Ref.~\cite{BL} to a supersymmetric model since it can also achieve that goal. 

In this letter we investigate the simplest supersymmetric extension of one of the models in Ref.~\cite{BL}. 
Unlike the nonsupersymmetric case here (if there are no large Fayet Illiopoulos D terms) the B and L symmetry breaking 
scales are necessarily of order the soft supersymmetry breaking scale. We discuss the main features of the model including 
the properties of the spectrum and dark matter candidates. We show that there are no dangerous operators that cause proton 
decay even after baryon and lepton number are spontaneously broken. This model should be interpreted as an effective theory 
below a scale that is at most a few orders of magnitude above the weak scale because beyond that point the Yukawa couplings 
of the fourth generation become strong~\cite{Godbole:2009sy}. Consequently the evidence for a supersymmetric extension 
of the standard model based on the meeting of the gauge couplings is not applicable in models with a fourth generation. 

Within the effective field theory approach it is possible to consider  gauge theories that are anomalous.  
With a  cutoff  that is only a few orders of magnitude above the weak scale it is possible to do this  in theories 
that gauge B and L~\cite{Preskill}. However, we prefer not to take that approach and cancel the anomalies in B and L using a fourth generation.

This paper is organized as follows: In section II we discuss baryon number violation in models where B and L are spontaneously broken.
The B and L violation in the MSSM is discussed in section III. In section IV we propose the simplest supersymmetric extension 
of the model in Ref.~\cite{BL}, while in section V we summarize our main findings.
\\
{\section{II. Baryon Number Violation in Models with B and L Spontaneously Broken Gauge Symmetries}}

Recently, we proposed  simple extensions of the Standard Model where B and L are local gauge symmetries~\cite{BL}.
These models are based on the gauge symmetry, $SU(3)_C \bigotimes SU(2)_L \bigotimes U(1)_Y \bigotimes U(1)_B \bigotimes U(1)_L$ 
and one introduces a new fermionic family to cancel all baryonic and leptonic anomalies. There are two ways to cancel all baryonic and leptonic anomalies, with  a new family of fermions that has the following properties. In Model I, one adds $Q_L^{'}$, $u_R^{'}$, $d_R^{'}$ with $B=-1$, and $l_L^{'}$, $e_R^{'}$, $\nu_R^{'}$ with $L=-3$, while in Model II the new generation has different chirality: $Q_R^{'}$, $u_L^{'}$, $d_L^{'}$, with $B=1$, and $l_R^{'}$, $e_L^{'}$ and $\nu_L^{'}$ with $L=3$. Since the new fourth generation fermions have different B and L quantum numbers than the quarks and leptons in the first three generations it was easy to arrange that are 
no flavour changing neutral currents at tree level. In order to avoid a stable fourth generation quark, we introduced a new 
scalar field which is a cold dark matter candidate that coupled the fourth generation fermions to first three generations. 
For a discussion  of the cosmology of these models including the generation of the baryon excess see ~\cite{Tim-Mark-Pavel}.
For generic studies of models with fourth generations see Ref.~\cite{4G}.

Since the fourth generation Yukawa couplings get strong at an energy scale not very far above the TeV scale these models have a fairly low ultraviolet cutoff and hence it is important that nucleon decay is forbidden even including non-renormalizable operators of very high dimension. 
In these models lepton number and baryon number are broken by the vacuum expectation value of fields $S_L$ and $S_B$ with L and B charges $n_L$ and $n_B$, respectively. In the calculation of S-matrix elements  lepton number and baryon number violation arises from insertions of the vacuum expectation values of these fields. Possible nucleon decay modes are: $p \rightarrow \pi^0 e^+$,  $p \rightarrow \pi^0 e^+\nu \nu$, $p \rightarrow  \pi^0e^+\nu {\bar \nu}$, {\it etc}. All possible nucleon decay modes have $\Delta{\rm B}=-1$ and $\Delta{\rm L}=\pm {\rm  ~an~ odd~natural~ number}$. Hence if $k |n_B| \ne  1$ and/or $k|n_L| \ne {\rm  ~an~ odd~natural~ number}$, for $k=1,2,\ldots$, proton decay is forbidden even allowing non renormalizable operators of arbitrarily high dimension. Clearly it is not difficult to arrange that baryon number violating nucleon decay is forbidden in models where baryon number and lepton number are gauged by a suitable choice of the charges $n_B$ and $n_L$ even though these symmetries are spontaneously broken. Note we are assuming here that the gravitino mass is greater than the proton mass. If it is lighter then final states without a lepton would be allowed.

Two body scattering process that violate baryon number can occur inside of the nucleus. 
For example $p+n \rightarrow \pi^+\pi^0$, $p+p\rightarrow  \pi^+ \pi^+, K^+ K^+$, {\it etc}. These along with $n-{\bar n}$ oscillations 
are forbidden if $k|n_B| \ne 2$. If they are not forbidden, by the value of $n_B$, the limits they impose on the scale 
of baryon number symmetry breaking are typically not  extremely strong because in the low energy effective theory (below the scales of spontaneous baryon number  and weak symmetry breaking) the lowest dimension operators that induce $\Delta{\rm B}=2$ transitions 
have six quark fields and are dimension nine. For a discussion of discrete symmetries that enforce baryon number 
and lepton number conservation in supersymmetric versions of the standard model see~\cite{Ibanez:1991pr}.

Models I and II in Ref.~\cite{BL} have several scalars with masses that are at or below the weak scale and this requires multiple 
fine tunnings ({\it \i.e.}, the hierarchy puzzle). Furthermore even though we assumed the breaking of B and L occurred at the 
weak scale there was no reason for this to be the case. Motivated by these issues  we study in this letter a simple supersymmetric 
extension of Model I. The quantum numbers of the quark and lepton fields are the same as Model I in~\cite{BL} but the scalar representations used to break the symmetry are different. Furthermore no additional scalars are introduced to prevent the stability of the fourth generation quarks. In the supersymmetric version of model I that we discuss below they decay through non-renormalizable interactions.  

\vskip0.3in

\section{III.  B and L violation in the MSSM }
The MSSM superpotential up to dimension five is given by
\begin{eqnarray}
{\cal W}_{MSSM}&=&  {\cal W}_{M} \ + \ {\cal W}_{L} \ + \ {\cal W}_{B} \ + {\cal W}_{5}.
\end{eqnarray}  
The first term in the superpotential,
\begin{eqnarray}
{\cal W}_{M}&=& g_u \hat{Q} \hat{u}^c \hat{H}_u \ + \  g_d \hat{Q} \hat{d}^c \hat{H}_d \ + \ g_e \hat{L} \hat{e}^c \hat{H}_d 
\nonumber \\
& + &  \mu \hat{H}_u \hat{H}_d, 
\end{eqnarray} 
contains all the renormalizable terms conserving matter parity, $M=(-1)^{3(B-L)}$. The terms violating L at the renormalizable level appear in
\begin{equation}
{\cal W}_{L} = \epsilon \hat{L} \hat{H}_u \ + \ \lambda \hat{L} \hat{L} \hat{e}^c \ + \ \lambda^{'} \hat{Q} \hat{L} \hat{d}^c.
\end{equation}
There is only one term in the MSSM superpotential which violates B at the renormalizable level and it is given by
\begin{equation}
{\cal W}_{B}= \lambda^{''}\hat{u}^c \hat{d}^c \hat{d}^c.
\end{equation}
Now, at the non-renormalizable level one also has the following dimension five operators that violate B and/or L\footnote{Note we 
have not yet gauged B and L.}:
\begin{eqnarray}
{\cal W}_{5}&=& \frac{\lambda_1}{\Lambda} \hat{Q} \hat{Q} \hat{Q} \hat{L} \ + \  
\frac{\lambda_2}{\Lambda} \hat{u}^c \hat{d}^c \hat{u}^c \hat{e}^c 
\ + \  \frac{\lambda_3}{\Lambda}  \hat{L} \hat{L} \hat{H}_u \hat{H}_u.
\end{eqnarray}
Using the interaction $\lambda^{'} \hat{Q} \hat{L} \hat{d}^c$ and the term in ${\cal W}_{B}$ one gets 
the dimension four contributions to proton decay, which predict a lifetime of  order $\tau_p \sim 10^{-15}$ years, if the couplings 
are order one and  the squark masses are around a 1 TeV. With similar assumptions the dimension five operators in ${\cal W}_{5}$ also give unacceptably fast contributions to the decay of the proton even if the $\Lambda$ scale is close to the Planck scale.
For a review on proton decay and a detailed discussion about these contributions see Ref.~\cite{review}. 

In order to clarify our notation we list the MSSM superfields:
\begin{eqnarray*}
\hat{Q} = \left(
\begin{array} {c}
\hat{u} \\ \hat{d}
\end{array}
\right) &\sim& {\small (3,2,1/6,1/3,0)}, 
\\
\hat{u}^c &\sim& {\small (\bar{3},1,-2/3,-1/3,0)},
\\  
\hat{d}^c &\sim& {\small (\bar{3},1,1/3,-1/3,0)},
\\
\hat{L} = \left(
\begin{array} {c}
 \hat{\nu} \\ \hat{e}
\end{array}
\right) &\sim& {\small (1,2,-1/2,0,1)},
\end{eqnarray*}
and
$
\hat{e}^c \sim {\small (1,1,1,0,-1)}. 
$
Notice that we have included their transformation properties under the gauge group 
$SU(3)_C \bigotimes SU(2)_L \bigotimes U(1)_Y \bigotimes U(1)_B \bigotimes U(1)_L$, 
anticipating that we will eventually gauge B and L.

The two MSSM Higgses are given by 
\begin{eqnarray*}
\hat{H}_u = \left(
\begin{array} {c}
	\hat{H}_u^+
\\
	\hat{H}_u^0
\end{array}
\right) \ \sim \ (1, 2, 1/2, 0, 0),
\\
\hat{H}_d = \left(
\begin{array} {c}
	\hat{H}_d^0
\\
	\hat{H}_d^-
\end{array}
\right) \ \sim \ (1, 2, -1/2, 0, 0).
\end{eqnarray*}
Adding right handed neutrinos, $\hat{\nu}^c \sim (1,1,0,0,-1)$, we have the following extra terms in the superpotential
\begin{eqnarray}
{\cal W}_{\nu}&=&g_\nu \hat{L} \hat{H}_u \hat{\nu}^c \ + \ M_{\nu} \hat{\nu}^c \hat{\nu}^c
\nonumber \\
&+& \frac{\lambda_4}{\Lambda} \hat{L} \hat{L} \hat{e}^c  \hat{\nu}^c \ + \  \frac{\lambda_5}{\Lambda} \hat{Q} \hat{L} \hat{d}^c \hat{\nu}^c 
\ + \  \frac{\lambda_6}{\Lambda} \hat{u}^c \hat{d}^c \hat{d}^c \hat{\nu}^c.
\end{eqnarray} 
It is well-known that adding three copies of right-handed neutrinos one can gauge B-L and 
the dimension four operators that violate baryon and/or lepton number in 
${\cal W}_{B}$  and ${\cal W}_{L}$ are not allowed. However, even  if 
we impose $B-L$ as a gauge symmetry the dimension five contributions 
to proton decay that arise from couplings in ${\cal W}_{5}$ are allowed. 
Therefore, one does not resolve the issue of an unacceptably large proton decay rate  in SUSY theories  just by gauging $B-L$. 
For a study of the origin of B and L violating interactions in B-L models see Ref~\cite{PRL}.
This is one of the main motivations to consider the SUSY version of the model proposed in Ref.~\cite{BL}.

In Ref.~\cite{Ko} the authors studied a supersymmetric extension of our model in Ref.~\cite{BL}. However, their motivation was primarily a study of dark matter candidates in the model while our motivation is to construct the simplest possible SUSY extensions of our model that do not permit proton decay even including non renormalizable terms of high dimension. We use nonrenormalizable interactions to render the fourth generation quarks unstable instead of adding additional multiplets as was done in~\cite{BL}.  
Note that stable color triplet heavy particles give rise to exotic nuclei that form atoms.  Limits on the density of such atoms and constraints from Big Bang nucleosynthesis suggest that stable heavy quarks with masses of a few hundred GeV are not acceptable.
\section{IV.  The MSSM  with B and L gauged}
In order to write the simplest supersymmetric model based on the gauge symmetry
\begin{center}
$G_{BL}=SU(3)_C \bigotimes SU(2)_L \bigotimes U(1)_Y \bigotimes U(1)_B \bigotimes U(1)_L$
\end{center}
and cancel anomalies we need to introduce chiral superfields for a new fermionic generation. They are:
\begin{eqnarray*}
\hat{Q}_4 = \left(
\begin{array} {c}
\hat{u}_4 \\ \hat{d}_4
\end{array}
\right) &\sim& {\small (3, 2, 1/6,-1, 0)}, 
\\ 
\hat{u}^c_4 &\sim& {\small (\bar{3},1,-2/3,1,0)},
\\
\hat{d}^c_4 &\sim& {\small (\bar{3},1,-1/3,1,0)},
\\
\hat{L}_4 = \left(
\begin{array} {c}
 \hat{\nu}_4 \\ \hat{e}_4
\end{array}
\right) &\sim& {\small (1,2,-1/2,0,-3)},
\\
\hat{e}^c_4 &\sim& {\small (1,1,1,0,3)},
\\
\hat{\nu}^c_4 &\sim& {\small (1,1,0,0,3)}.
\end{eqnarray*}

We have shown explicitly how the new fermions transform under  $G_{BL}$. We need additional chiral superfields that acquire vacuum expectation values that break
B and L. The required new Higgses to break $U(1)_B$ are: $\hat{S}_B \sim (1,1,0,-1/3,0)$ and $\hat{\bar{S}}_B \sim (1,1,0,1/3,0)$.  For the chiral superfields that break $U(1)_L$ there are two possibilities that we consider: either  (i) $\hat{S}_{L} \sim (1,1,0,0,-6)$ and $\hat{\bar{S}}_{L} \sim (1,1,0,0,6)$ or (ii) $\hat{S}_{L} \sim (1,1,0,0,-2)$ and $\hat{\bar{S}}_{L} \sim (1,1,0,0,2)$.

The superpotential of the theory is given by
\begin{eqnarray}
{\cal W}_{BL}&=& {\cal W}_{\rm{Yukawa}} \ + \  {\cal W}_{\rm{Higgs}} \ + \ {\cal W}^5_{BL},
\end{eqnarray}
where in case (i),
\begin{eqnarray}
&&{\cal W}^{(i)}_{\rm{Yukawa}}= g_u \hat{Q} \hat{u}^c \hat{H}_u \ + \  g_d \hat{Q} \hat{d}^c \hat{H}_d \ + \ g_e \hat{L} \hat{e}^c \hat{H}_d \nonumber 
\\
&& \ + \  g_\nu \hat{L} \hat{H}_u \hat{\nu}^c  \ + \ Y_{u} \hat{Q}_4 \  \hat{H}_u \ \hat{u}_4^c \ + \  Y_{d} \hat{Q}_4 \  \hat{H}_d \ \hat{d}_4^c 
\nonumber \\
& & \ + \  Y_{e} \hat{L}_4 \  \hat{H}_d \ \hat{e}_4^c \ + \ Y_{\nu} \hat{L}_4 \  \hat{H}_u \ \hat{\nu}_4^c \ + \  \lambda_{\nu^c_4}  \hat{\nu}^c_4 \hat{\nu}^c_4 \hat{S}_L.
\end{eqnarray}
Here the ordinary three generation neutrinos have Dirac masses and the fourth generation neutrino has both Dirac and Majorana mass terms.  The fourth generation neutrino mass must be greater than $M_Z/2$. On the other hand
for case (ii)
\begin{eqnarray}
&&{\cal W}^{(ii)}_{\rm{Yukawa}}=g_u \hat{Q} \hat{u}^c \hat{H}_u \ + \  g_d \hat{Q} \hat{d}^c \hat{H}_d \ + \ g_e \hat{L} \hat{e}^c \hat{H}_d \nonumber 
\\
&& \ + \ g_\nu \hat{L} \hat{H}_u \hat{\nu}^c  \ + \ Y_{u} \hat{Q}_4 \  \hat{H}_u \ \hat{u}_4^c \ + \  Y_{d} \hat{Q}_4 \  \hat{H}_d \ \hat{d}_4^c 
\nonumber \\
& & \ + \  Y_{e} \hat{L}_4 \  \hat{H}_d \ \hat{e}_4^c \ + \ Y_{\nu} \hat{L}_4 \  \hat{H}_u \ \hat{\nu}_4^c \ + \  \lambda_{\nu^c}  \hat{\nu}^c \hat{\nu}^c \hat{\bar{S}}_L  \nonumber \\
 && \ + \ \lambda_{\nu^c}^{'}  \hat{\nu}^c \hat{\nu}^c_4 \hat{S}_L.
\end{eqnarray}
The ordinary light three generations of neutrinos have both Majorana and Dirac mass terms and so extremely small Yukawa coupling constants can be avoided using the type I see-saw mechanism~\cite{TypeI}. The fourth generation neutrino has a Dirac mass term and a Majorana mass term that mixes it with the first three generations of neutrinos. 

The Higgs part of the superpotential is
\begin{eqnarray}
{\cal W}_{\rm{Higgs}}&=& \mu \hat{H}_u \hat{H}_d \ + \ \mu_B \hat{S}_B \hat{\bar{S}}_B \ + \ \mu_L \hat{S}_L \hat{\bar{S}}_L. 
\end{eqnarray}
Finally the dimension five terms that allow fourth generation particles to decay to the ordinary generations are
\begin{eqnarray}
{\cal W}^5_{BL} &=&  
 \frac{a_1}{\Lambda} \hat{u}^c_4 \hat{d}^c \hat{d}^c \hat{S}_B 
\ + \    \frac{a_2}{\Lambda} \hat{u}^c \hat{d}^c_4 \hat{d}^c \hat{S}_B \ + \ 
\frac{a_3}{\Lambda} \hat{\nu}^c \hat{\nu}^c \hat{\nu}^c \hat{\nu}^c_4. 
\nonumber \\
\end{eqnarray}
The terms proportional to the $a_1$ and $a_2$ couplings are needed to avoid a stable quark from the $4$th generation.  
In case (i) the term proportional to $a_3$ avoids the presence of a stable heavy Dirac neutrino. Notice that here we write 
only the relevant dimension five operators.

For simplicity in our discussions we ignore kinetic mixing between the $U(1)$'s and the possible Fayet-Illiopoulos D-terms. 

\underline{Symmetry Breaking:}
Here we investigate the symmetry breaking mechanism to show that $U(1)_B$ and $U(1)_L$ can be 
broken at the TeV scale. In the case of the $U(1)_B$ symmetry, it is broken by the vev of the scalar fields, ${S}_B$ and 
${\bar{S}}_B$. These vacuum expectation values can be chosen real and positive.
The relevant soft terms for our discussion are:
\begin{eqnarray}
- \Delta{\cal L}_{Soft} &= & \left( - \ b_{B} S_B \bar{S}_B \ + \  \text{h.c.} \right) 
\nonumber \\
& + &  m_{S_B}^2 |S_B|^2 \ + \ m_{\bar{S}_B}^2 |\bar{S}_B|^2,
\end{eqnarray}
For simplicity of notation we take $b_B$ to be real. Using $\left<S_B \right>=v_{B}/\sqrt{2}$ 
and $\left<\bar{S}_B\right>=\bar{v}_B/\sqrt{2}$ for the vevs one finds
\begin{eqnarray}
V_B &=& \frac{1}{2}|\mu_{B}|^2 \left( v^2_B + \bar{v}^2_B \right) \ - \ b_{B} v_B \bar{v}_B \ + \ \frac{1}{2}m_{S_B}^2 v^2_B 
\nonumber \\
& + & \frac{1}{2} m_{\bar{S}_B}^2 \bar{v}^2_B  \ + \  \frac{g_{B}^2}{32} n_{B}^2 \left( v^2_B - \bar{v}^2_B \right)^2.
\end{eqnarray}
Now, assuming that the potential is bounded from bellow along the D-flat direction we get:
\begin{equation}
2 b_{B} < 2 |\mu_{B}|^2 + m_{S_B}^2 + m_{\bar{S}_B}^2.
\end{equation} 
while
\begin{equation}
b_{B}^2 > \left( |\mu_{B}|^2 + m_{S_B}^2 \right) \left( |\mu_{B}|^2 + m_{\bar{S}_B}^2 \right),
\end{equation}
in order to have a non-trivial vacuum. Minimizing with respect $v_B$ and $\bar{v}_B$ one finds that
\begin{eqnarray}
|\mu_{B}|^2 + m_{S_B}^2 - \frac{1}{2} M_{Z_{B}}^2 \cos 2 \beta_B - b_{B} \cot \beta_B=0,
\\
|\mu_{B}|^2 + m_{\bar{S}_B}^2 + \frac{1}{2} M_{Z_{B}}^2 \cos 2 \beta_B - b_{B} \tan \beta_B=0,
\end{eqnarray}
with $\tan \beta_B=v_B/\bar{v}_B$ and $M_{Z_{B}}^2=(n_B g_{B})^2 (v^2_B + \bar{v}^2_B)/4$.
Here $n_B=1/3 (-1/3)$ for $\bar{S}_B (S_B)$. The above equations can be written as
\begin{eqnarray}
\frac{1}{2} m_{Z_B}^2 &=& -|\mu_B|^2 \ - \ \left(  \frac{m_{S_B}^2 \tan^2 \beta_B - m_{\bar{S}_B}^2}{\tan^2 \beta_B - 1}\right),
\\
b_B &=& \frac{\sin 2 \beta_B}{2} \left(  2 |\mu_B|^2 + m_{S_B}^2 + m_{\bar{S}_B}^2\right).
\end{eqnarray}
For symmetry breaking to occur  $m_{Z_B}^2 $ must be positive and it is clear from the above equation that the gauge boson mass is set by 
the soft supersymmetry breaking terms since they must overpower the negative contribution from the $\mu_B$ piece.
The $U(1)_B$ symmetry is broken at the SUSY scale. 

The analysis of $U(1)_L$ breaking is similar to the breaking of $B-L$ studied in Ref.~\cite{Fate}.
Several fields can get a VEV: $\left< S_L\right>$, $\left< \bar{S}_L \right>$, $\left< \tilde{\nu}\right>$ and $\left< \tilde{\nu}^c \right>$.
There are  two different cases:  i)  R-parity conservation, where only $S_L$ and $\bar{S}_L$ can get a VEV, 
and  ii) R-parity is spontaneously broken due to the VEV of sneutrinos. In the latter case one needs a tackyonic mass term~\cite{Fate} for 
the ``right-handed" sneutrinos. 
In this paper we assume that the soft supersymmetry breaking mass terms for the sneutrinos are not tackyonic  so that the only fields with lepton number that get a VEV are  $S_L$ and $\bar{S}_L$.
In the case where the sneutrinos get a vev, one has R-parity and L violating interactions, which together with the interactions coming from Eq.(20)
give rise to proton decay. Since the cutoff in the theory  is low due to the existence of the Landau poles for the fourth generation Yukawa couplings, 
one finds that these contributions give rise to unacceptably fast proton decay. For a study where the sneutrino vev breaks the leptonic symmetry see Ref.~\cite{PRL}.

In this paper we do not address the $\mu$ problem. The supersymmetric parameters $\mu$, $\mu_B$ and $\mu_L$ are taken to be of order the
supersymmetry breaking scale, even though there is no clear reason for this to be the case.

\underline{Baryon Number  Violation:} 
One does not generate operators that mediate proton decay because $S_L$ has an even lepton number charge (see section III). 
In the MSSM typically we define matter parity as $M=(-1)^{3(B-L)}=M_L \times M_B$, where $M_{L}=(-1)^{-3L}$ and 
$M_{B}=(-1)^{3B}$ can be called leptonic parity and baryonic parity, respectively. Notice that $M_L=-1$ for all leptons 
and $+1$ for $\hat{S}_L$ and $\hat{\bar{S}}_L$. All the fields with baryon number have $M_{B}=-1$. 
After symmetry breaking,  $M_L$ is conserved but $M_B$ is broken. The fact that $M_L$ is conserved tells us 
that one cannot generate any operator which induces proton decay, because one must break $M_L$ to allow 
the proton to decay.  Note that the absence of proton decay is true even if we include nonrenormalizable operators of 
arbitrarily high dimension. One can, however, generate $|\Delta {\rm B}|=2$ operators that mediate nucleus decay. 
For example, a dimension seven operator in the superpotential
\begin{equation}
\Delta{\cal W}^7_B= {{ \tilde{\lambda}^{''}} \over \Lambda^3}{\hat u}^c{\hat d}^c{\hat d}^c {\hat {\bar S}}_B^3 
\end{equation}
generates a contribution to the reaction~$^{16}O(pp)\to^{14}C K^+ K^+$ after integrating out the squarks and the gluino.
The relevant dimension nine operator is $C_9 \ \overline{u^c} \overline{d^c} \overline{s^c} \  \overline{u^c} \overline{d^c} 
\overline{s^c}$, with $C_9$:
\begin{equation}
C_9= \left( \frac{\tilde{\lambda}^{''}_{uds}\ v_B^3}{\Lambda^3} \right)^2 \times \frac{4 \pi \alpha_s}{M_{\tilde{s}^c}^4 M_{\tilde{g}}}.
\end{equation}  
Assuming that $M_{\tilde{s}^c}, M_{\tilde{g}} \sim 1$ TeV, and using the experimental limit on this channel from 
the Super-Kamiokande collaboration one finds that ${ \tilde \lambda^{''}}_{uds} v_B^3/ \Lambda^3 < 10^{-8}$~\cite{Goity,Thesis}. 
Notice that  $C_9$ can induce $n-\bar{n}$ oscillation at tree level if one assumes flavour violation in the squark sector. Here, 
for simplicity we do not consider this possibility. At one loop level, one has a contribution to $n-{\bar n}$ oscillations  where inside the loops 
one has the charginos (winos) and the SM quarks. However, constraints from $n -{\bar n}$ oscillations are weaker than the one from dinucleon decays discussed above.
For a review on $n-\bar{n}$ oscillation see Ref.~\cite{Mohapatra}.

The couplings above allow the squarks to decay to two quarks with a partial width of order $\Gamma ({\tilde q} \to q {\bar q}) \sim ({\tilde \lambda^{''}})^2(v_B/\Lambda)^6/(64 \pi)$. This of course means that (apart from the gravitino in models with a high enough scale of spontaneous supersymmetry breaking), the lightest neutralino is not a dark matter candidate\footnote{Another well motivated dark matter candidate is the axion since it is associated with the solution of the strong CP puzzle.}. It decays through a virtual squark to three light quarks. Baryon number violating neutralino decay was discussed in Ref.~\cite{Butterworth:2009qa}.   In this model the fourth generation squarks decay to quark pairs which also violates baryon number.

{\textit{\underline{Gauge Bosons}}}: Neglecting kinetic gauge boson mixing, in this theory we have a leptophobic $Z_B$ and quarkphobic $Z_L$ neutral gauge bosons 
associated to the new symmetries $U(1)_B$ and $U(1)_L$, respectively.   For a review on $Z^{'}$ models see Ref.~\cite{Zp}.
The masses of the new neutral gauge bosons are given by
\begin{eqnarray}
m_{Z_B} &=&\frac{g_B}{6} \left( v_B^2 \ + \  \bar{v}_{B}^2 \right)^{1/2}, \\
m_{Z_L} &=&\frac{n_L}{2} g_L \left( v_L^2 \ + \  \bar{v}_{L}^2 \right)^{1/2},
\end{eqnarray}
where in  case (i) $n_L=6$ and in case (ii) $n_L=2$. The collider constraints on a quark-phobic $Z^{'}$ are more severe than the case of $Z_B$.
For the case of $Z_L$ one can use the LEP2 bounds~\cite{PDG}, while for $Z_B$ it is possible to use the UA2 bounds~\cite{ZB}.

{\textit{\underline{Neutralinos}}}: 
The neutralino sector now has B and L neutralinos in addition to the MSSM neutralinos. In total one has the MSSM neutralinos
$\tilde{\chi}^0_i$, the baryonic neutralinos, $\tilde{\chi}^0_B=(\tilde{B}_B, \tilde{S}_B, \tilde{\bar{S}}_B)$. Here
$\tilde{B}_B$ is the $U(1)_B$ gaugino, and the $\tilde{S}_B$ Higgsinos. Finally, one also has the leptonic neutralinos,  
$\tilde{\chi}^0_L=(\tilde{B}_L, \tilde{S}_L, \tilde{\bar{S}}_L)$. Here $\tilde{B}_L$ is the $U(1)_L$ gaugino, 
and $\tilde{S}_L$ and $\tilde{\bar{S}}_L$ are the superpartners of the Higgses breaking the local leptonic symmetry. 
It is straightforward to work out the neutralino mass matrices. For example, the neutralino $\tilde{\chi}^0_B$ mass matrix  is,
\begin{equation}
	\mathcal{M}_{\tilde{\chi}^0_B} = 
	\left(
	\begin{array}{ccc}
		m_{B}
		&
		- \frac{g_B v_B}{6}  
		&
		\frac{g_B \bar{v}_B}{6}
	\\
		- \frac{g_B v_B}{6} 
		&
		0
		&
		-\mu_B
	\\
		\frac{g_B \bar{v}_B}{6} 
		&
		-\mu_B
		&
		0
	\end{array} \right),
\end{equation}
where $m_B$ is the bino mass, and $\mu_B$ is the mass term of the Higgsinos in the baryonic sector.
Notice that only when the Higgsino term is small one can have a light neutralino in this sector. 

\textit{\underline{Sfermions and New Higgs Spectrum}}: After symmetry breaking the sfermion masses get an 
extra contribution due to the new $D$-terms for $U(1)_L$ and $U(1)_B$. See Ref.~\cite{Maike} for a similar 
study of the spectrum of sfermions of a $U(1)$ extension of the MSSM. Of course we have additional
sfermions associated with the fourth generation. In order to illustrate this point we show the charged MSSM slepton masses
\begin{eqnarray}
M_{\tilde{e}_{L_i}}^2 &=& m_{\tilde{L}_i}^2 + m_{e_i}^2 - \left( \frac{1}{2} -  \sin^2 \theta_W \right) M_{Z}^2 \cos 2 \beta + D_L,
\nonumber 
\\
&&
\\
M_{\tilde{e}_{i}^c}^2 &=& m_{\tilde{e}^c_i}^2 + m_{e_i}^2 -  M_{Z}^2 \sin^2 \theta_W \cos 2 \beta - D_L,
\end{eqnarray}
with
\begin{eqnarray}
D_L &=& \frac{1}{2 n_L} m_{Z_L}^2 \cos 2 \beta_L.
\end{eqnarray}  
Here $m_{\tilde{L}}$ and $m_{\tilde{e}^c}$ are the soft terms for left and right-handed sleptons, respectively.
The new angle $\beta_L$ is defined as $\tan \beta_L = v_L / \bar{v}_L$.

There are  three new neutral L-Higgses: two CP-even $S_{L_1}, S_{L_2}$ and one CP-odd $A_L$, 
while in the $U(1)_B$-sector one has the neutral Higgses $S_{B_1}, S_{B_2}$ and $A_B$. 
These two sectors are not coupled to the MSSM sector at tree level through renormalizable interactions. For a recent study of the Higgs decays in 
the MSSM with four generations see Ref.~\cite{4MSSM}.

The masses of the Higgses in the baryonic sector are
\begin{equation}
m_{S_{B_1}, S_{B_2}}^2=\frac{1}{2} \left( m_{A_B}^2 + m_{Z_{B}}^2 
\mp \sqrt{D} \right),
\end{equation}
with
\begin{equation}
D=(m_{A_B}^2 - m_{Z_{B}}^2)^2  +  4 m_{Z_{B}}^2 m_{A_B}^2 \sin^2 (2 \beta_B) ,
\end{equation}
where
\begin{equation}
m_{A_B}^2=\frac{2 b_B}{\sin 2 \beta_B}.
\end{equation}
Notice that the Higgses in this sector can light because the limit on the mass of $Z_B$ is not very strong~\cite{ZB}.
In this way we conclude the discussion of the properties of the spectrum of our model.

\section{V. Summary and Outlook}
In this paper we have proposed  a simple model with baryon and lepton number gauged 
and spontaneously broken at the supersymmetry breaking scale. After symmetry breaking 
the leptonic matter parity is conserved and so proton decay is forbidden (provided the gravitino is heavier than the proton) 
even when nonrenormalizable operators of arbitrarily high dimension are included. 

We have noted some of the important features associated with the spontaneous breaking of baryon number 
including the implications for dark matter candidates. We have pointed out some properties of the spectrum and 
possible baryon number violating decays. It is important to mention that in this model one can have 
lepton number violating signals from the decays of the right-handed neutrinos and baryon number violating 
signals from the decays of squarks and gauginos without conflict with the bounds coming from proton decay, 
$n-\bar{n}$ oscillations and dinucleon decays. It would be interesting to investigate the collider signals and 
cosmological aspects of this model including the possibility of weak scale baryogenesis. 

\vskip0.3in

{\textit{Acknowledgments}}:
P.F.P. would like to thank the Institute for Advanced Study in Princeton for hospitality in the last part of this project.
P.F.P. thanks S. Spinner for several discussions. The work of  P.F.P. has been supported in part by the U.S. Department of
Energy under grant No. DE-FG02-95ER40896, and by the Wisconsin
Alumni Research Foundation. The work of M.B.W. was supported in part by the 
U.S. Department of Energy under contract No. DE-FG02-92ER40701.



\begin{thebibliography}{000}
\bibitem{Foot:1989ts}
  S.~Rajpoot,
  ``Electroweak Interactions With Gauged Baryon And Lepton Numbers,''
  Phys.\ Rev.\  {\bf D40 } (1989)  2421;
  X.~-G.~He, S.~Rajpoot,
  ``Anomaly Free Left-right Symmetric Models With Gauged Baryon And Lepton Numbers,''
  Phys.\ Rev.\  {\bf D41 } (1990)  1636;
  R.~Foot, G.~C.~Joshi and H.~Lew,
  ``Gauged baryon and lepton numbers,''
  Phys.\ Rev.\  D {\bf 40}, 2487 (1989).

\bibitem{Carone}
  C.~D.~Carone and H.~Murayama,
  ``Realistic models with a light U(1) gauge boson coupled to baryon number,''
  Phys.\ Rev.\  D {\bf 52}, 484 (1995)
  [arXiv:hep-ph/9501220].
  
\bibitem{BL}
  P.~Fileviez P\'erez and M.~B.~Wise,
  ``Baryon and Lepton Number as Local Gauge Symmetries,''
  Phys.\ Rev.\  D {\bf 82} (2010) 011901
  [Erratum-ibid.\  D {\bf 82} (2010) 079901]
  [arXiv:1002.1754 [hep-ph]].


\bibitem{Manuel}
M. Drees, R. Godbole and P. Roy, {\textit{Theory and Phenomenology of Sparticles}}, 
(World Scientific, 2004).

\bibitem{Lee:2010hf}
See for example:
  K.~S.~Babu, I.~Gogoladze and K.~Wang,
  ``Gauged baryon parity and nucleon stability,''
  Phys.\ Lett.\  B {\bf 570} (2003) 32
  [arXiv:hep-ph/0306003];
  H.~S.~Lee and E.~Ma,
  ``Gauged ${\rm B}-\xi {\rm L}$ origin of R parity and its implications,''
  Phys.\ Lett.\  B {\bf 688}, 319 (2010)
  [arXiv:1001.0768 [hep-ph]].
  
  \bibitem{Preskill}
  J.~Preskill,
  ``Gauge anomalies in an effective field theory,''
  Annals Phys.\  {\bf 210}, 323 (1991).
  
  \bibitem{Godbole:2009sy}
  R.~M.~Godbole, S.~K.~Vempati, A.~Wingerter,
  ``Four Generations: SUSY and SUSY Breaking,''
  JHEP {\bf 1003}, 023 (2010).
  [arXiv:0911.1882 [hep-ph]].

 
\bibitem{Tim-Mark-Pavel}
  T.~R.~Dulaney, P.~Fileviez Perez, M.~B.~Wise,
  ``Dark Matter, Baryon Asymmetry, and Spontaneous B and L Breaking,''
  Phys.\ Rev.\  {\bf D83 } (2011)  023520.
  [arXiv:1005.0617 [hep-ph]].  
  
\bibitem{4G}
See for example:
  P.~H.~Frampton, P.~Q.~Hung and M.~Sher,
  ``Quarks and leptons beyond the third generation,''
  Phys.\ Rept.\  {\bf 330} (2000) 263
  [arXiv:hep-ph/9903387];
  B.~Holdom, W.~S.~Hou, T.~Hurth, M.~L.~Mangano, S.~Sultansoy and G.~Unel,
  ``Four Statements about the Fourth Generation,''
  PMC Phys.\  A {\bf 3} (2009) 4
  [arXiv:0904.4698 [hep-ph]];
  G.~D.~Kribs, T.~Plehn, M.~Spannowsky and T.~M.~P.~Tait,
  ``Four generations and Higgs physics,''
  Phys.\ Rev.\  D {\bf 76} (2007) 075016
  [arXiv:0706.3718 [hep-ph]];
  L.~M.~Carpenter,
  ``Fourth Generation Lepton Sectors with Stable Majorana Neutrinos: From LEP
  to LHC,''
  arXiv:1010.5502 [hep-ph];
  L.~M.~Carpenter and A.~Rajaraman,
  ``Revisiting Constraints on Fourth Generation Neutrino Masses,''
  Phys.\ Rev.\  D {\bf 82} (2010) 114019
  [arXiv:1005.0628 [hep-ph]];
  A.~J.~Buras, B.~Duling, T.~Feldmann, T.~Heidsieck and C.~Promberger,
  ``Lepton Flavour Violation in the Presence of a Fourth Generation of Quarks
  and Leptons,''
  JHEP {\bf 1009} (2010) 104
  [arXiv:1006.5356 [hep-ph]];
  A.~J.~Buras, B.~Duling, T.~Feldmann, T.~Heidsieck, C.~Promberger and S.~Recksiegel,
  ``Patterns of Flavour Violation in the Presence of a Fourth Generation of
  Quarks and Leptons,''
  JHEP {\bf 1009} (2010) 106
  [arXiv:1002.2126 [hep-ph]];
  S.~Dawson and P.~Jaiswal,
  ``Four Generations, Higgs Physics, and the MSSM,''
  Phys.\ Rev.\  D {\bf 82} (2010) 073017
  [arXiv:1009.1099 [hep-ph]];
  W.~Y.~Keung and P.~Schwaller,
  ``Long Lived Fourth Generation and the Higgs,''
  arXiv:1103.3765 [hep-ph].
  
\bibitem{Ibanez:1991pr}
  L.~E.~Ibanez, G.~G.~Ross,
  ``Discrete gauge symmetries and the origin of baryon and lepton number conservation in supersymmetric versions of the standard model,''
  Nucl.\ Phys.\  {\bf B368}, 3-37 (1992);
  H.~K.~Dreiner, C.~Luhn and M.~Thormeier,
  ``What is the discrete gauge symmetry of the MSSM?,''
  Phys.\ Rev.\  D {\bf 73} (2006) 075007
  [arXiv:hep-ph/0512163].

\bibitem{review}
  P.~Nath and P.~Fileviez P\'erez,
  ``Proton stability in grand unified theories, in strings, and in branes,''
  Phys.\ Rept.\  {\bf 441} (2007) 191;
 [arXiv:hep-ph/0601023].

\bibitem{PRL}
P.~Fileviez P\'erez and S.~Spinner,
  ``Spontaneous R-Parity Breaking and Left-Right Symmetry,''
  Phys.\ Lett.\  B {\bf 673} (2009) 251;
 [arXiv:0811.3424 [hep-ph]].
  V.~Barger, P.~Fileviez Perez, S.~Spinner,
  ``Minimal gauged U(1)(B-L) model with spontaneous R-parity violation,''
  Phys.\ Rev.\ Lett.\  {\bf 102 } (2009)  181802.
  [arXiv:0812.3661 [hep-ph]].

\bibitem{Ko}
  P.~Ko and Y.~Omura,
  ``Supersymmetric $U(1)_B \times U(1)_L$ model with leptophilic and leptophobic cold
  dark matters,''
  arXiv:1012.4679 [hep-ph].
  
 \bibitem{TypeI}
  P.~Minkowski,
  ``Mu $\to$ E Gamma At A Rate Of One Out Of 1-Billion Muon Decays?,''
  Phys.\ Lett.\ B {\bf 67} (1977) 421;
  T. Yanagida,
in {\it Proceedings of the Workshop on the Unified Theory
   and the Baryon Number in the Universe}, eds. O. Sawada et al.,
p.~95, KEK Report 79-18, Tsukuba (1979);
  M. Gell-Mann, P. Ramond and R. Slansky,
   in {\it Supergravity}, eds. P. van Nieuwenhuizen et al.,
   (North-Holland, 1979), p.~315;
  S.L. Glashow, in {\it Quarks and Leptons}, Carg\`ese, eds. M. L\'evy et al.,
(Plenum, 1980), p. 707;
  R.~N.~Mohapatra and G.~Senjanovi\'c,
  ``Neutrino Mass And Spontaneous Parity Nonconservation,''
  Phys.\ Rev.\ Lett.\  {\bf 44} (1980) 912.


\bibitem{Fate}
  P.~Fileviez Perez and S.~Spinner,
  ``The Fate of R-Parity,''
  Phys.\ Rev.\  D {\bf 83} (2011) 035004
  [arXiv:1005.4930 [hep-ph]].
  
\bibitem{Goity}
  J.~L.~Goity and M.~Sher,
  ``Bounds on delta B = 1 couplings in the supersymmetric standard model,''
  Phys.\ Lett.\  B {\bf 346} (1995) 69
  [Erratum-ibid.\  B {\bf 385} (1996) 500]
  [arXiv:hep-ph/9412208];

\bibitem{Thesis}
Michael D. Litos, ``A search for dinucleon decay into kaons using the SK water cherenkov detector",
Ph.D. Thesis, Boston University, 2010.

\bibitem{Mohapatra}
  R.~N.~Mohapatra,
  ``Neutron-Anti-Neutron Oscillation: Theory and Phenomenology,''
  J.\ Phys.\ G {\bf 36} (2009) 104006
  [arXiv:0902.0834 [hep-ph]].
  
  \bibitem{Butterworth:2009qa}
  J.~M.~Butterworth, J.~R.~Ellis, A.~R.~Raklev, G.~P.~Salam,
  ``Discovering baryon-number violating neutralino decays at the LHC,''
  Phys.\ Rev.\ Lett.\  {\bf 103}, 241803 (2009).
  [arXiv:0906.0728 [hep-ph]].
  
\bibitem{Zp}
  P.~Langacker,
  ``The Physics of Heavy Z' Gauge Bosons,''
  arXiv:0801.1345 [hep-ph].
  
\bibitem{PDG}
  C.~Amsler {\it et al.}  [Particle Data Group],
  ``Review of Particle Physics,''
  Phys.\ Lett.\  B {\bf 667} (2008) 1.


\bibitem{ZB}
  M.~R.~Buckley, D.~Hooper, J.~Kopp and E.~Neil,
  ``Light Z' Bosons at the Tevatron,''
  arXiv:1103.6035 [hep-ph];
  M.~Buckley, P.~Fileviez Perez, D.~Hooper and E.~Neil,
  ``Dark Forces At The Tevatron,''
  arXiv:1104.3145 [hep-ph];
  V.~D.~Barger, K.~M.~Cheung and P.~Langacker,
  ``Baryonic Z-prime connection of LEP R(b,c) data with Tevatron (W, Z, gamma)
  b anti-b events,''
  Phys.\ Lett.\  B {\bf 381} (1996) 226
  [arXiv:hep-ph/9604298];
  F.~Yu,
  ``A Z' Model for the CDF Dijet Anomaly,''
  arXiv:1104.0243 [hep-ph];
  K.~Cheung and J.~Song,
  ``Tevatron Wjj Anomaly and the baryonic $Z'$ solution,''
  arXiv:1104.1375 [hep-ph].

\bibitem{Maike}
  P.~Fileviez Perez, S.~Spinner and M.~K.~Trenkel,
  ``The LSP Stability and New Higgs Signals at the LHC,''
  arXiv:1103.5504 [hep-ph]; 
  ``Testing the Mechanism for the LSP Stability at the LHC,''
  arXiv:1103.3824 [hep-ph].
  
\bibitem{4MSSM}
  R.~C.~Cotta, J.~L.~Hewett, A.~Ismail, M.~P.~Le and T.~G.~Rizzo,
  ``Higgs Properties in the Fourth Generation MSSM: Boosted Signals Over the 3G
  Plan,''
  arXiv:1105.0039 [hep-ph].
 

\end{thebibliography}
\end{document}